\documentclass[aps,pra,reprint,floatfix,showpacs,twocolumn]{revtex4-1}
\usepackage[utf8]{inputenc}
\usepackage{amsfonts}
\usepackage{amssymb}
\usepackage{amsmath}
\usepackage{amstext}
\usepackage{graphicx}
\usepackage{wasysym}

\begin{document}
\title{An extension of Bogoliubov theory for a many-body system with a time scale hierarchy: the quantum mechanics of second Josephson oscillations}

\author{M.\,P.\,Strzys}
\email{strzys@physik.uni-kl.de}
\author{J.\,R.\,Anglin}
\affiliation{OPTIMAS Research Center and Fachbereich Physik, Technische Universit\"at Kaiserslautern, D--67653 Kaiserslautern, Germany}


\begin{abstract}
Adiabatic approximations are a powerful tool for simplifying nonlinear quantum dynamics, and are applicable whenever a system exhibits a hierarchy of  time scales. Current interest in small nonlinear quantum systems, such as few-mode Bose-Hubbard models, warrants further development of adiabatic methods in the particular context of these models. Here we extend our recent work on a simple four-mode Bose-Hubbard model with two distinct dynamical time scales, in which we showed that among the perturbations around excited stationary states of the system is a slow collective excitation that is not present in the Bogoliubov spectrum. We characterized this mode as a resonant energy exchange with its frequency shifted by nonlinear effects, and referred to it as a second Josephson oscillation, in analogy with the second sound mode of liquid helium II. We now generalize our previous theory beyond the mean field regime, and construct a general Bogoliubov free quasiparticle theory that explicitly respects the system's adiabatic invariant as well the exact conservation of particles. We compare this theory to the numerically exact quantum energy spectrum with up to forty particles, and find good agreement over a significant range of parameter space.
\end{abstract}

\maketitle

\section{Introduction}

Interest in the nonlinear quantum dynamics of small systems has grown in recent years, as theory and experiments together approach the mesoscopic regime. With Bose-Einstein condensates (BEC) in optical lattices now available in many laboratories, few-mode Bose-Hubbard (BH) systems provide a realization that is both clean and rich \cite{Bloc05,Gati07,Myat97,Matt99,Nemo00,Vard01b,Sakm10,Chia11b,Gill11}. From a fundamental point of view, small, isolated quantum systems have long served as testbeds for studying equilibration and thermalization \cite{Berm05,Berm04,Gira60,Kino06,Rigo07}. Recent theoretical work has sought to reformulate the basic concepts of equilibrium \cite{Deut91,Sred94,Rigo08,Rigo09a} and equilibration \cite{Pono11a,Pono11b}. The essential challenge remains great, however: Even for quite small quantum systems, the complete energy spectra can be computationally unattainable. Such standard approximations as Bogoliubov linearization normally only yield good results in the neighborhood of the ground state, but a full understanding of thermal effects must encompass excited states as well. 

\begin{figure}[htp]
\begin{center}
 \includegraphics[width = 0.48\textwidth]{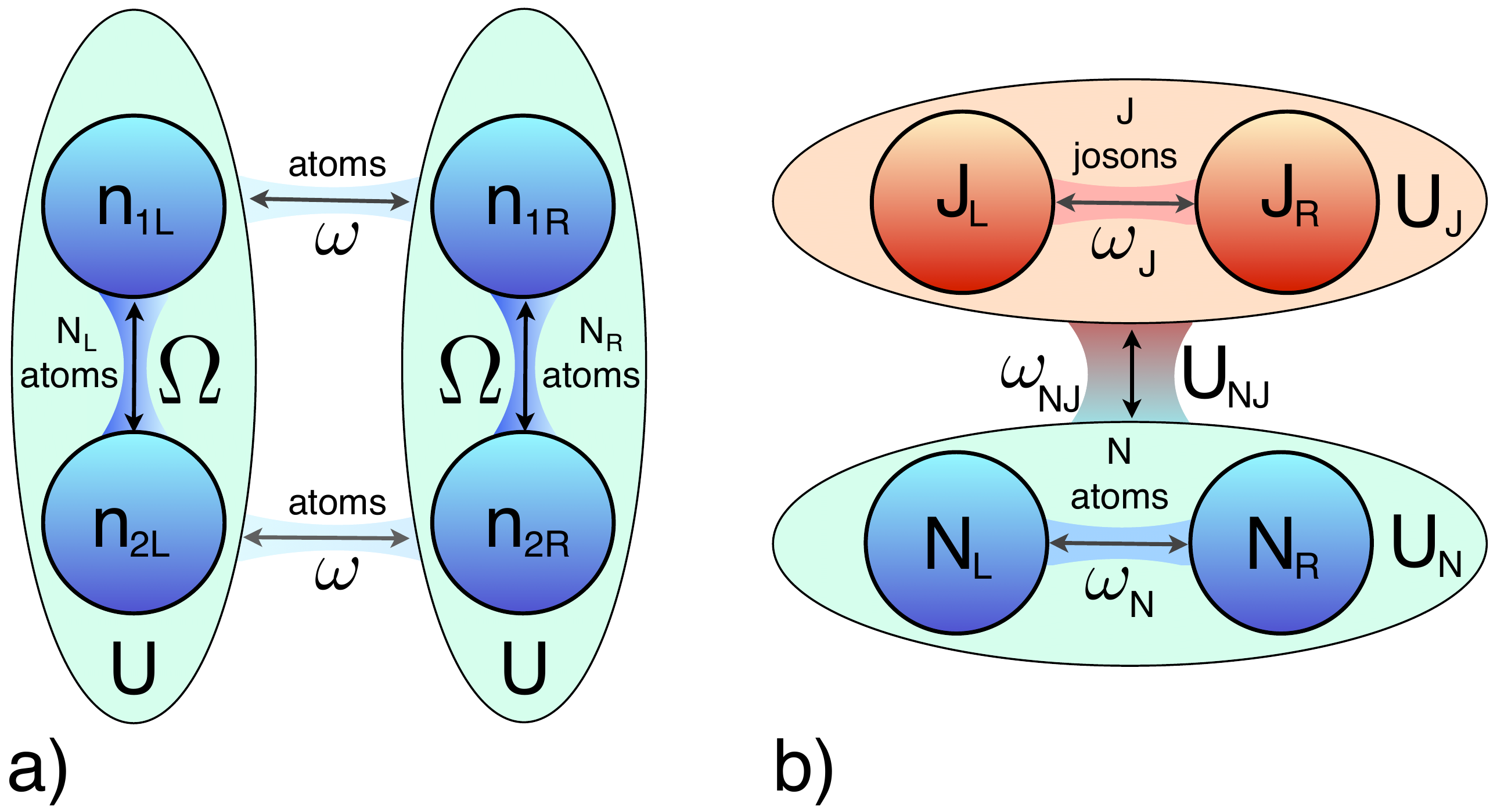}
 \caption{(Color online) Two ways of looking at the same four-mode Bose-Hubbard system: a) $N$ conserved atoms occupying four modes perform multiple Josephson oscillations with a distinct time scale hierarchy $\omega / \Omega \ll 1$; b) In an adiabatic sense, the total number $J$ of high-frequency Josephson excitations (`josons') is also conserved.  Hence we may also regard the full system as consisting of $N$ atoms and $J$ josons, all interacting nonlinearly within two weakly-coupled subsystems.}
\label{sysfig}
\end{center}
\end{figure}

Adiabatic methods may provide an additional theoretical tool for understanding BH systems, especially since it is experimentally possible to tune tunneling rates over large ranges in order to impose a time scale hierarchy. 
In our previous work \cite{Strz10} we introduced a system of two of such two-mode BH subsystems coupled as to represent a four-mode BH system with a large time-scale separation. Each subsystem in this model is a pair of bosonic modes coupled (for example, by tunneling between two wells) and thus represents an idealized Josephson junction \cite{Milb97,Ragh99,Giov00,Levy07,Truj09}. The time-scale separation is implemented by a clear hierarchy of the tunneling frequencies coupling the four modes, \textit{i.e.}~the intra subsystem coupling $\Omega$ is chosen to be by far larger than the inter subsystem coupling $\omega$; cf Fig.~\ref{sysfig}a. 
Therefore the subsystem characteristic timescale -- the Josephson frequency for each of the more tightly coupled mode pairs -- is to be the shortest one in the entire system. In \cite{Strz10} we showed that, while the total particle number $N$ is exactly conserved, in adiabatic theory there also exists another conserved quantity, the total number $J$ of Bogoliubov excitations of both of the two-mode BH subsystems. Since these are the elementary excitations of idealized Josephson junctions, we referred to them as `josons'. Just like atoms, josons can be distributed arbitrarily between the two subsystems, but their total number is conserved [see Fig.~\ref{sysfig}b]. Thus the subsystems may exchange both atoms, by ordinary (`first') Josephson oscillations, and josons (\textit{i.e.}, high-frequency energy), in a second low frequency collective mode that we termed `second Josephson oscillations', by analogy with first and second sound in helium II. 

Whereas in \cite{Strz10} we confirmed our adiabatic results with comparisons to numerical Gross-Pitaevskii mean field theory, in this paper we now explicitly address the quantum energy spectrum of our four-mode system, and show how even highly excited levels can be given accurately by our generalization of standard number-conserving Bogoliubov theory. We then drop our previous restriction to the mean field regime of large $N$, and relax the restriction $J \ll N$. In the spirit of our previous work we are then able to derive an extended Bogoliubov theory which treats atoms and joson quasiparticles similarly. Implemented via Holstein-Primakoff transformations (HPT) \cite{Hols40}, this formalism ensures the conservation of the total number of atoms $N$ as well as the total number of josons $J$, and is thus both technically and qualitatively somewhat different from what is usually called number-conserving Bogoliubov theory \cite{Cast98,Gard97,Gard07,Oles08}.  Standard Bogoliubov theory only works well in the neighborhood of the $N$-particle ground state of the system, but our nonlinearly resummed version covers the neighborhood of the ground state of each energy band for given $N$ and $J$. This validity range includes many highly excited states of the full four-mode system.

The extensions we will report here provide a useful counterpart to other recent work on this same rich four-mode system. In \cite{Chia11}, a truncated Wigner approach was used for dynamical calculations. Although the so-called truncated Wigner approximation only represents non-classical dynamics in a limited sense, since it uses strictly classical (Liouville) evolution for the Wigner function, the results in \cite{Chia11} should be accurate in the mean field regime, and otherwise give at least some indication of quantum effects. Although \cite{Chia11} attempts to test our prediction of second Josephson oscillations, the specific cases reported there either have $J$ essentially zero, so that second Josephson oscillations do not occur because there are no josons present, or else begin with very large differences in the populations of the four modes, so that no linear theory can be expected to perform well, and so that their Fock state simulations may be sampling many trajectories close to unstable classical fixed points, for which mean field theory breaks down unusually quickly. In this sense \cite{Chia11} and our earlier paper are complementary rather than conflicting, and this paper is complementary to both, in providing fully quantum mechanical results for small perturbations around many excited but stationary states.

This paper will be arranged as follows. In the first section we will give a short review of standard Bogoliubov theory for the simple case of a bosonic two-mode system. In the following section we will derive the extended version of our theory for larger values of $J$. The third section will be devoted to the structure of the quantum spectrum and reveals the double ladder structure of the energy bands. The last section will focus on the dependence of the levels on the interactions, present the numerical results, and discuss the breakdown of the free quasiparticle theory. We will close with a general discussion.

\section{A review of Bogoliubov theory in the two-mode system}\label{section:twomode}
                                                                                                                                                                                                                                                                                                                           
\begin{figure}[htp]
\begin{center}
 \includegraphics[width = 0.3\textwidth]{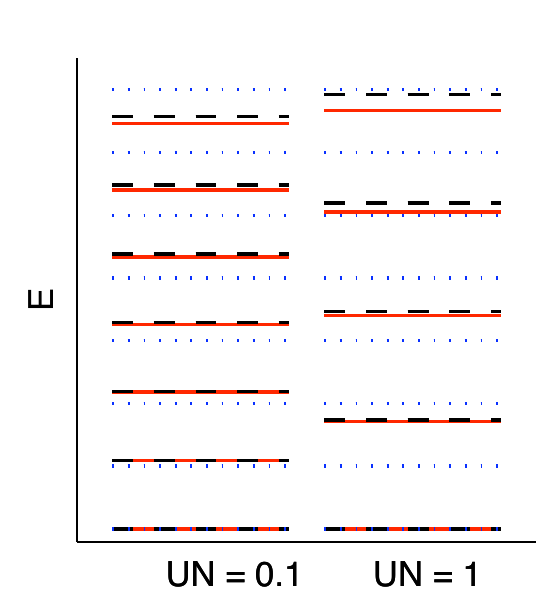}
 \caption{(Color online) Comparison of the different approximations to the exact two-mode spectrum for $\Omega = 1$ and $N=30$: solid red line: exact levels; dotted blue line: free particles; dashed black line: Bogoliubov free quasiparticles, all in units of $\hslash \Omega$}
\label{linspec2}
\end{center}
\end{figure}

The simplest case of the BH model is its two-mode version. This simple model has been very extensively studied, since it exhibits nontrivial effects such as self-trapping, admits a thorough analytic treatment \cite{Milb97,Smer97,Ragh99,Vard01b} and is also experimentally accessible in cold atom laboratories \cite{Bloc05,Gati07,Myat97,Matt99}. Here it shall serve as a simple and well-known example to recall the technique of standard (symmetry-breaking) Bogoliubov theory \cite{Bogo47,Genn66,Bogo67,Fett72}. The two-mode BH Hamiltonian reads as
\begin{equation}\label{twomodeH}
 \hat{H}_2 =  -\frac{\Omega}{2}\left(\hat a_{1}^{\dagger}\hat a_{2} +\hat a_{2}^{\dagger}\hat a_{1} \right) + \frac{U}{2}\sum_{i=1}^{2} \hat a_{i}^{\dagger 2}\hat a_{i}^{2}
\end{equation}
where $\hat a_i,\hat  a_i^\dagger$ are the annihilation and creation operators of the two bosonic modes. A first approximation to such a system would be to neglect the interactions completely, \textit{i.e.}~to set $U \rightarrow 0$, and thereby obtain equally spaced energy levels (the quanta of Rabi oscillations). However, since the interaction energy scales quadratically with particle number, this approximation can only be accurate for $U$ being very small compared to the coupling $\Omega$, which however is indeed true in many experimental realizations. Nevertheless there exists a much broader regime in which the energy levels do remain linearly spaced, to a very good approximation; but their spacing can be substantially different from that given by the simple free particle theory with $U=0$. This frequency shift depends on the interactions and increases with growing $U$. Bogoliubov theory is able to predict these corrections correctly and thus rather accurately describes the excitations above the ground state even for higher interactions, even though it is only a linearization. In Fig.~\ref{linspec2} the exact energy levels of the Hamiltonian \eqref{twomodeH} are compared to the free particle and the Bogoliubov approximation. 
To obtain the Bogoliubov Hamiltonian one diagonalizes the linear term in \eqref{twomodeH} and assumes the symmetric mode to be highly occupied, in the sense that its annihilation and creation operators can be replaced by the number $\sqrt{N}$, to yield an effective single-mode theory. Keeping only Hamiltonian terms up to quadratic order in $\hat{a}$ and $\hat{a}^{\dagger}$, one diagonalizes these with a Bogoliubov transformation \cite{Hols40,Bogo47} by introducing new bosonic quasiparticle annihilation and creation operators $\hat b,\hat b^\dagger$ such that $\hat{a} = u\hat{b} + v\hat{b}^\dagger$, where $u^2-v^2=1$ to maintain canonical commutation relations. This procedure leads to a very simple Hamiltonian of the form
\begin{equation}\label{twomodeHbog}
 \hat H_{\rm Bog} = E_{0}+\sqrt{\Omega (\Omega + UN)}\; \hat b^\dagger \hat b
\end{equation}
where $E_{0}$ is an irrelevant $c$-number constant. Here and throughout this paper we assume units such that $\hslash = 1$. The Bogoliubov approximation can therefore be understood as a theory of free quasiparticles, where the Rabi frequency $\Omega$ of the two-mode system is replaced by the Josephson frequency $\tilde \Omega = \sqrt{\Omega (\Omega + UN)}$ (which of course reduces to $\Omega$ in the limit $U\to0$). The $U$-dependence of $\tilde \Omega$ is the crucial feature for describing the low-energy excitations accurately in the presence of interactions. Thus Bogoliubov theory is a noninteracting quasiparticle theory that includes nonlinear interactions among the real particles in a mean field regime with one strongly occupied mode. 

There are several ways to improve the naively classical treatment of the highly occupied mode, without significantly altering Bogoliubov theory otherwise, by using the exact number conservation of the system. In this paper as well as in our previous work, we have implemented number conservation rigorously and exactly by using Holstein-Primakoff transformations \cite{Strz10}, which is slightly different from the standard formalism introduced in \cite{Cast98,Gard97,Gard07,Oles08}.

It is important to appreciate that Bogoliubov transformations are simply a class of linear canonical transformations, in the sense of Hamiltonian mechanics, and that there is no intrinsic restriction on the kind of excitations to which Bogoliubov theory can be applied. In particular it is possible to perform multiple Bogoliubov transformations in succession, defining a revised set of higher order quasiparticles to absorb certain interactions among the quasiparticles that were defined by the first Bogoliubov transformation. As we shall see in the following paragraphs it is possible in this fashion to construct a generalized number-conserving Bogoliubov approach that is able to handle excitations not only above the ground state, but above a whole class of states, namely the lowest energy states of various energy bands. This extended Bogoliuov theory is therefore able to cover a far larger part of the spectrum.

\section{Extended joson theory}

The four-mode BH system we consider consists of two coupled two-mode systems and is described by a Hamiltonian of the form 
\begin{eqnarray}\label{Hamiltonian}
 \hat H &=& \hat{H}_{L}+\hat{H}_{R} + \hat{H}_{LR},\\
 \hat{H}_{\alpha} &=&  -\frac{\Omega}{2}\left(\hat a_{1\alpha}^{\dagger}\hat a_{2\alpha} +\hat a_{2\alpha}^{\dagger}\hat a_{1\alpha} \right) + U\sum_{i=1}^{2} \hat a_{i\alpha}^{\dagger 2}\hat a_{i\alpha}^{2},\nonumber\\
 \hat{H}_{LR} &=& -\frac{\omega}{2}\left(\hat a_{1L}^{\dagger}\hat a_{1R} + \hat a_{2L}^{\dagger}\hat a_{2R} + \textrm{H. c.}\right)\nonumber
\end{eqnarray}
with on-site interaction $U$ and index $\alpha=L,R$ for the left and right subsystems, respectively. A clear time scale hierarchy is implemented by  choosing $\Omega$, the high tunneling rate within each subsystem pair, to be far larger than the low rate $\omega$ between the two subsystems, \textit{i.e.}~restricting to the limit $\omega / \Omega \ll 1$. However, as we will see in our numerical analysis, this is not actually a very stringent condition; the adiabatic results remain remarkably accurate as long as some degree of time scale separation is present. It would not require very much optimism to replace the $\ll1$ condition with $\lesssim1$.

\subsection{Uncoupled subystems}

We will begin by considering uncoupled two-mode subsystems and introduce atom-moving operators $\hat{a}_{\alpha},\hat{a}_{\alpha}^{\dagger}$ via an HPT:
\begin{eqnarray}\label{schwing}
	\sqrt{\hat{N}_{\alpha}-\hat{a}^{\dagger}_{\alpha}\hat{a}_{\alpha}}\ \hat{a}_{\alpha} \equiv \frac{1}{2}(\hat{a}_{1\alpha}^{\dagger}+\hat{a}_{2\alpha}^{\dagger})(\hat{a}_{1\alpha}-\hat{a}_{2\alpha})\;.
\end{eqnarray}
These new single-index operators commute with $\hat{N}_{\alpha}=\sum_{i=1,2}\hat{a}_{i\alpha}^{\dagger}\hat{a}_{i\alpha}$ and fulfill the usual bosonic commutation relations $[\hat{a}_{\alpha},\hat{a}^{\dagger}_{\alpha}]=1$. Note that since $\hat{a}_{\alpha}$ transfers one of the $\hat{N}_{\alpha}$ atoms, only states with 
\begin{equation}\label{bedingung}
\hat{a}^{\dagger}_{\alpha}\hat{a}_{\alpha} \leqslant \hat{N}_{\alpha}
\end{equation}
(\textit{i.e.}~only superpositions of eigenstates of these operators whose eigenvalues satisfy this inequality) can be physical.
With these definitions the Hamiltonians of the subsystems assume the form
\begin{eqnarray}\label{Halpha}
 \hat{H}_{\alpha} &=&  -\frac{\Omega}{2}\hat{N}_\alpha + \Omega \hat{a}_{\alpha}^{\dagger}\hat{a}_{\alpha} + \frac{U}{2}\hat{N}_\alpha(\hat{N}_\alpha-2)\\
 &+& \frac{U}{2}\left(\hat{a}^\dagger_\alpha\sqrt{\hat{N}_\alpha-\hat{a}^\dagger_\alpha \hat{a}_\alpha} + \sqrt{\hat{N}_\alpha-\hat{a}^\dagger_\alpha \hat{a}_\alpha}\hat{a}_\alpha\right)^2\nonumber.
\end{eqnarray}
It will be observed that this Hamiltonian provides no coupling between the physical and unphysical Hilbert spaces, and so the implicit introduction of unphysical states with $\langle\hat{a}^{\dagger}_{\alpha}\hat{a}_{\alpha}\rangle > \langle\hat{N}_{\alpha}\rangle$ has done no actual harm.

So far we have followed exactly the procedure of \cite{Strz10}. To now go beyond our previously assumed limit $J/N \ll 1$, in this paper we must take some care in treating the last term in \eqref{Halpha}. In our previous approach we expanded this term in inverse powers of $\sqrt{N_\alpha}$ and omitted terms of the order of $U N_\alpha^{-1}$. But this in fact is a rather bold approximation, as will become clear if we do all the transformations at once. With the help of a second HPT we introduce a number-conserving operator $\hat{a}$ transferring atoms between the $L$ and $R$ subsystems, which satisfies $[\hat{a},\hat{a}^{\dagger}]=1$ and commutes with the total particle number $\hat{N} = \hat{N}_{L}+\hat{N}_{R}$ and the $\hat{a}_\alpha$ modes. For the particle number on either side we then get the expression
\begin{eqnarray}
 \hat{N}_{L,R} &\equiv& \frac{1}{2}\left[ N \pm (\hat a^\dagger \sqrt{N-\hat a^\dagger \hat a} + \sqrt{N-\hat a^\dagger \hat a}\,\hat a)\right]. 
 \end{eqnarray}
On the $\hat a_\alpha$ modes we at first perform a Bogoliubov transformation \cite{Hols40,Bogo47} letting $\hat a_\alpha = u\hat b_\alpha + v\hat b_\alpha^\dagger$ with $[\hat{b}_{\alpha},\hat{b}^{\dagger}_{\alpha}]=1$, where we leave $u$ and $v$ arbitrary at the moment. These parameters will be computed later self-consistently. Since we demand the adiabaticity constraint $\omega \ll \Omega$ and are only interested in dynamical frequencies on the order of $\omega$, we may apply a rotating wave approximation (RWA) discarding all terms not commuting with the number of Bogoliubov excitations, \textit{i.e.}~josons, $\hat b_\alpha^\dagger \hat b_\alpha$ on each side. These are simply the elementary excitations of the two idealized Josephson junctions. In this regime the total joson number 
\begin{equation}
\hat{J} = \hat{b}^{\dagger}_{L}\hat{b}_{L}+\hat{b}^{\dagger}_{R}\hat{b}_{R}
\end{equation}
is an adiabatic invariant and may be treated as a new constant of motion. This motivates us to finally define a joson-moving operator $\hat b$, by one last HPT, according to
\begin{eqnarray}\label{HPT3}
	\sqrt{J-\hat{b}^{\dagger}\hat{b}}\ \hat{b} \equiv \frac{1}{2}(\hat{b}_{L}^{\dagger}+\hat{b}_{R}^{\dagger})(\hat{b}_{L}-\hat{b}_{R})\,,
\end{eqnarray}
completely analogous to \eqref{schwing}. As a direct consequence of \eqref{bedingung} we get the restriction $ J \leqslant N $ for physical states, since josons are carried by atoms after all, as they simply reflect different atom distributions.
We may then express $\hat a_\alpha^\dagger \hat a_\alpha$ in term of the newly defined modes $\hat a$ and $\hat b$, yielding
\begin{equation}
 \hat a_\alpha^\dagger \hat a_\alpha = \frac{u^2+v^2}{2}\left[J \pm (\hat b^\dagger \sqrt{J-\hat b^\dagger \hat b} + \sqrt{J-\hat b^\dagger \hat b}\,\hat b) \right] + v^2.
\end{equation}
From this it is obvious that for larger $J$ the last term of \eqref{Halpha} may more accurately be expanded in inverse powers of $\sqrt{N-\tilde J}$, where
\begin{equation}
\tilde J = (u^2+v^2)J
\end{equation}
is the scaled joson number. So far, however, the Bogoliubov parameters have been involved in the transformations, but have not been properly defined yet. This has to be done now, self-consistently, by solving a third order equation for $u^2 + v^2$,
\begin{eqnarray}
 \Omega(\Omega + UN) + \frac{U^2 N^2}{4} =  -\Omega U J &&(u^2+v^2)^3 \nonumber\\
+ \left( \Omega (\Omega + UN)- \frac{U^2 J^2}{4} \right) &&(u^2+v^2)^2 \\
+ \left( \Omega UJ + \frac{U^2 NJ}{4} \right)&&(u^2+v^2)\;, \nonumber
\end{eqnarray}
while keeping the normalization condition $u^2 - v^2=1$. For compactness in the following we will continue to write $u$ and $v$ as parameters, but one should keep in mind that in fact they turn out to be functions of $\Omega$, $U$, $N$ and $J$. Thus, the leading term of the Hamiltonian is just of the standard Bogoliubov type, \textit{i.e.} \begin{equation}
 \hat H_L + \hat H_R = \tilde\Omega J + \mathcal{O}(U \Omega^0)
\end{equation}
with the slightly modified Bogoliubov frequency 
\begin{equation}\label{bogfrec}
 \tilde \Omega = \sqrt{\Omega \left(\Omega + U(N-\tilde J)\right)}
\end{equation}
which also has to be determined self-consistently. Note that it is not identical with the standard Bogoliubov frequency of the two-mode system in \eqref{twomodeHbog}, but is shifted in the presence of josons. Being interested in the low-frequency dynamics only, we ignore the constant parts of the Hamiltonian and up to fourth order in the $\hat a_\alpha$ modes finally obtain in terms of atom- and joson-moving operators
\begin{eqnarray}\label{HLHR}
 \hat H_L &+& \hat H_R = \frac{U}{4}(\hat a^\dagger \sqrt{N-\hat a^\dagger \hat a} + \sqrt{N-\hat a^\dagger \hat a}\,\hat a)^2\\
 &-& \frac{U}{2}(u+v)^2 (u^2+v^2)C_{1}(\hat b^\dagger \sqrt{J-\hat b^\dagger \hat b} + \sqrt{J-\hat b^\dagger \hat b}\,\hat b)^2 \nonumber\\ 
 &+& \frac{U}{2}(u^2+v^2)C_{1}(\hat a^\dagger \sqrt{N-\hat a^\dagger \hat a} + \sqrt{N-\hat a^\dagger \hat a}\,\hat a)\nonumber\\
 &&\times (\hat b^\dagger \sqrt{J-\hat b^\dagger \hat b} + \sqrt{J-\hat b^\dagger \hat b}\,\hat b)\nonumber
\end{eqnarray}
where the constant $C_{1} = 1 - \mathcal{O}(v^4 / (N-\tilde J)^2)$. It is to be pointed out that the last term in \eqref{HLHR} is an atom-joson coupling. This coupling may be considered as a lingering reminder of the fact that, although josons and atoms can be considered as independent conserved particles in the low-frequency regime, fundamentally josons are motional excitations that must be carried by atoms. The atom-joson coupling is thus a matter of definition as much as of dynamics.

\subsection{Turning on the coupling}

Let us now turn to the coupling term $H_{LR}$ of the total Hamiltonian \eqref{Hamiltonian}, and perform the transformations of the preceding section once again. Here once more terms occur that need careful treatment. $(1-\hat a_\alpha^\dagger \hat a_\alpha / \hat N_\alpha)^{1/2}$, for instance, must be expanded in powers of $\sqrt{N-\tilde J}$, as discussed above. To be consistent we expand again up to fourth order in the $\hat a_\alpha$ modes. In RWA we then obtain a coupling Hamiltonian of the form
\begin{eqnarray}\label{Hnonlin}
 \hat H_{LR} &=& \omega C_2 \hat a^\dagger \hat a + \omega (u^2+v^2)\hat b^\dagger \hat b \\
 &-& \frac{\omega}{2}(u^2+v^2) C_3 (\hat a^\dagger \sqrt{N -\hat a^\dagger \hat a} + \sqrt{N -\hat a^\dagger \hat a}\,\hat a)\nonumber\\
 &&\times(\hat b^\dagger \sqrt{J -\hat b^\dagger \hat b} + \sqrt{J -\hat b^\dagger \hat b}\,\hat b) \nonumber\\
 &-& \frac{\omega}{2N}(u^2+v^2)(\hat a^\dagger \sqrt{N -\hat a^\dagger \hat a} - \sqrt{N -\hat a^\dagger \hat a}\,\hat a)\nonumber\\
 &&\times(\hat b^\dagger \sqrt{J -\hat b^\dagger \hat b} - \sqrt{J -\hat b^\dagger \hat b}\,\hat b) \nonumber\\
 &+& \frac{\omega}{4}(C_3 - C_2/N) (\hat a^\dagger \sqrt{N -\hat a^\dagger \hat a} + \sqrt{N -\hat a^\dagger \hat a}\,\hat a)^2\nonumber\\
 &+& \frac{\omega}{4}(u^2+v^2) C_3 (\hat b^\dagger \sqrt{J -\hat b^\dagger \hat b} + \sqrt{J -\hat b^\dagger \hat b}\,\hat b)^2 \nonumber
\end{eqnarray}
where we introduced the constants
\begin{eqnarray}
 C_2 &=& 1 - \mathcal{O}(v^2/N),\\
 C_3 &=& \frac{1}{N-\tilde J} + \mathcal{O}(v^2/(N-\tilde J)^2).\nonumber
\end{eqnarray}
Thus we have computed the whole four-mode Hamiltonian in terms of atom- and joson-moving operators $\hat a$ and $\hat b$. Note that in contrast to our derivation in \cite{Strz10} there appear terms which vanish in the limit $J/N\to0$, for example those proportional to $\sqrt{J/N}$. These terms are necessary for ensuring that the whole transformation performed on the original operators is canonical; if they are omitted in the small $J$ limit, then the operators only fulfill the canonical commutation relations up to corrections of the order of $1/N$. This observation does not invalidate our derivation in \cite{Strz10}, since there we consistently omitted terms of that order, and hence could also omit the sub-leading terms in $J/N$. By taking these terms into account, now we may also treat states with large values of $J \approx N$.

To now identify the collective excitations we may simply linearize \eqref{Hnonlin} in $\hat a$ and $\hat b$ which yields
\begin{eqnarray}\label{Hlin}
 \hat{H}_{\textrm{lin}} &=& \omega_N \hat a^\dagger\hat a + \omega_J \hat b^\dagger\hat b + \frac{\omega_{NJ}}{2}(\hat a^\dagger - \hat a)(\hat b^\dagger - \hat b)\nonumber\\
 &+& \frac{U_N}{4} N(\hat a^\dagger + \hat a)^2 + \frac{U_J}{4} J (\hat b^\dagger + \hat b)^2\\
 &+& \frac{U_{NJ}}{2}\sqrt{NJ}(\hat a^\dagger + \hat a)(\hat b^\dagger + \hat b)\nonumber 
\end{eqnarray}
where we used the abbreviations
\begin{eqnarray}\label{abbrev}
 \omega_N &=& \omega C_2, \quad \omega_J = \omega(u^2+v^2),\quad \omega_{NJ} = -\omega\sqrt{\frac{J}{N}},\nonumber\\
 U_N &=& U + \omega(C_3-C_2/N)\\
 U_J &=& \omega(u^2+v^2)C_3 - 2U (u+v)^2(u^2+v^2)C_1, \nonumber\\
 U_{NJ} &=& U(u+v)^2 C_1 - \omega(u^2+v^2)C_3.\nonumber
\end{eqnarray}
The Hamiltonian \eqref{Hlin} now just represents a coupled linear two-mode system which can immediately be diagonalized by another Bogoliubov transformation of the form
\begin{equation}
\label{bog2}
\binom{\hat a}{\hat b} = \mathbf{U} \binom{\hat a'}{\hat b'} + \mathbf{V}\binom{\hat a'^\dagger}{\hat b'^\dagger}
\end{equation}
where the matrices with the Bogoliubov parameters have to satisfy the standard constraints $\mathbf{U}\mathbf{U}^\dagger - \mathbf{V}\mathbf{V}^\dagger = \mathbf{1}$ and $\mathbf{U}\mathbf{V}^\mathrm{T} = \mathbf{V}\mathbf{U}^\mathrm{T}$ to ensure canonical bosonic commutation relations $[\hat a', \hat a'^\dagger] = 1$ and $[\hat b', \hat b'^\dagger] = 1$ for the new operators. Therefore the Hamiltonian \eqref{Hlin} can be written in the simple form 
\begin{equation}\label{twomodes}
 \hat{H}_{\textrm{lin}} = \tilde\omega_+ \hat a'^\dagger \hat a' + \tilde\omega_- \hat b'^\dagger \hat b'  = \tilde\omega_+ \hat n + \tilde\omega_- \hat j 
\end{equation}
with two independent quantum numbers $n, j$. The decoupled collective modes, \textit{i.e.}~the eigenfrequencies of \eqref{Hlin}, may easily be determined using the Heisenberg equations of motion 
$\mathrm{i}\dot{\hat a}' = [\hat a',\hat H_\mathrm{lin}] = \tilde \omega_+ \hat a'$ and $\mathrm{i}\dot{\hat b}' = [\hat b',\hat H_\mathrm{lin}] = \tilde \omega_- \hat b'$ respectively and turn out to be
\begin{eqnarray}\label{finalmodes}
 \tilde{\omega}_\pm^{2} &=& -\omega_{NJ} U_{NJ} \sqrt{NJ} + \frac{ \tilde{\omega}_N^2+ \tilde{\omega}_{J}^2}{2} \pm \left[\left(\frac{ \tilde  \omega_N^2 -  \tilde \omega_J^2}{2}\right)^2 \right.\nonumber\\
  &-& \omega_{NJ} U_{NJ} \sqrt{NJ} (\tilde \omega_N^2+ \tilde \omega_J^2) + \omega_N\omega_J U_{NJ}^2 NJ\nonumber\\ 
 &+& \left.\omega_{NJ}^2(\omega_J + U_J J)(\omega_N + U_N N)\phantom{\frac{1}{1}}\right]^{1/2}\;.
\end{eqnarray}
Here we introduced the atom Josephson frequency
\begin{eqnarray}
 \tilde \omega_N &=& \sqrt{\omega_N(\omega_N + U_N N)}\\ 
  &=& \sqrt{\omega(\omega + UN)} + \mathcal{O}(\omega U)\;,\nonumber
\end{eqnarray}
which is equal to the lowest mode of the standard Bogoliubow formalism up to corrections of the order of $\omega U$, and the new Josephson frequency for josons
\begin{equation}
 \tilde \omega_J = \sqrt{\omega_J(\omega_J + U_J J)}\;.
\end{equation}
For the rest of the paper we will refer to this linearized effective theory \eqref{twomodes} as extended Bogoliubov theory, since it is a quasiparticle theory that includes interactions among the josons in a similar way as standard Bogoliubov theory includes interactions among the atoms.

\section{The spectrum}

Since the coupling terms in \eqref{Hlin} are rather small, the mode-mixing due to the Bogoliubov transformation \eqref{bog2} is also small, and to a good approximation $n$ counts the atomic excitations, while $j$ counts the joson excitations. 
Thus, in this adiabatic linear limit the spectrum of the full four-mode quantum system is organized as follows. The energies are grouped into bands with fixed quantum numbers $N$ and $J$. Since the joson excitations result from an unequal distribution of the atoms over the modes, not all $(n,j)$ combinations are physical, because $J\leqslant N$ has to be fulfilled. Each of the bands can be described by a Hamiltonian of the form \eqref{twomodes} and therefore assumes a `pyramidal' structure due to the range of allowed values of $n$ and $j$.

To get the numerical results for the full four-mode quantum systems, we numerically diagonalize the Hamiltonian. The dimension of the Hilbert space is $\textrm{dim}(\mathcal{H}) = (N+3)(N+2)(N+1)/6$; desktop computing power has allowed us to handle total particle numbers up to $40$. The pyramidal band structure predicted by the adiabatic approximation is immediately recognizable as a close approximation to the exact spectrum, to the point where we can unambiguously label each exact eigenstate with the quantum numbers $(N,J,n,j)$ of its adiabatic approximation. The energy eigenstates are of course all simultaneously eigenstates of  $\hat{N}$. The adiabatic invariance of $J$ is reflected in their having expectation values of $\hat{J}$ that are all very close to integers, with very small quantum uncertainties. For each exact energy eigenstate, therefore, we can identify its $(N,J)$ band according to the adiabatic theory by defining the exact state's $J$ quantum number to be the rounding of the exact expectation value $\langle\hat{J}\rangle$ to the nearest integer. The states within each band can be assigned their internal band quantum numbers $\left|n,j\right>$ in similar fashion, by rounding the nearly-integral expectation values of the corresponding operators.

\begin{figure}[htp]
\begin{center}
 \includegraphics[width = 0.4\textwidth]{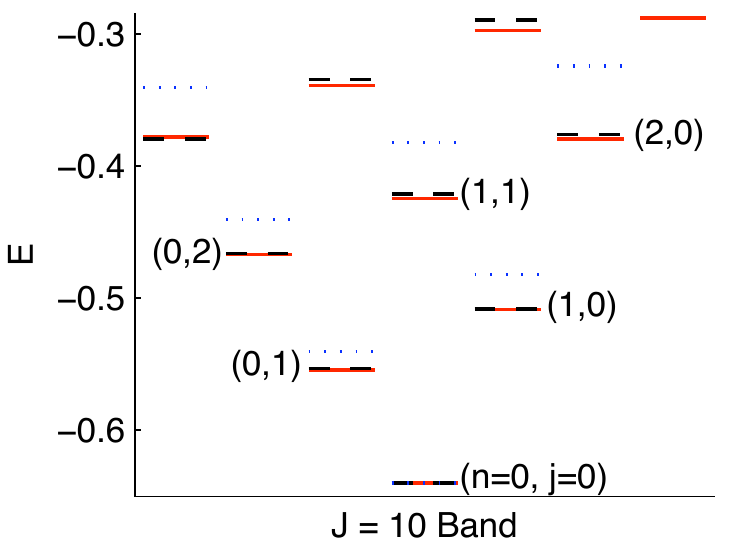}
 \caption{(Color online) Comparison of the different approximations to the exact four-mode spectrum (for $N=10$, $J=3$, $\Omega=1$, $\omega=0.1$ and $UN = 0.05$): solid red line: exact levels; dashed black line: extended Bogoliubov; dotted blue line: standard Bogoliubov free quasiparticles, all in units of $\hslash \Omega$}
\label{pyramid}
\end{center}
\end{figure}

To quantitatively analyze the mode frequencies, let us now turn to the spectrum of the full four-mode system itself. This is in principle exactly similar to the analysis for the two-mode case presented in Sec. \ref{section:twomode}. We may now compare the numerically exact spectrum to the free quasiparticle spectrum from the standard Bogoliubov approach, and our extended Bogoliubov theory, which includes interactions among the Bogoliubov quasiparticles. Since for fixed $N$ the spectrum is organized in $J$-bands, we may restrict ourselves to one of these, and group the states according to their $n,j$-excitations. The result for $N=10$, $\langle\hat{J}\rangle\doteq 3$ and the moderate interaction of $UN = 0.05$ is shown in Fig.~\ref{pyramid}. 
The `pyramidal' structure which appears correctly resembles the linear Hamiltonian with two quantum numbers. This structure is truncated, however, since for $N, J$ fixed both $n$ and $j$ are bounded from above, as they must fulfill $0 \leqslant n \leqslant N-J$ and $0 \leqslant j \leqslant J$, respectively. (This restriction is of no practical consequence, since the extended Bogoliubov theory breaks down for these highly excited states anyway. Nevertheless it is possible to derive a complementary theory, quite analogously to this one, which correctly describes the upper end of each $J$-band, leaving poorly described only the middle ranges of large bands.) It can be seen that for low excitations within any band the extended Bogoliubov theory is a clear improvement over the standard Bogoliubov approximation.

\section{Focus on interactions}

\begin{figure}[htp]
\begin{center}
 \includegraphics[width = 0.4\textwidth]{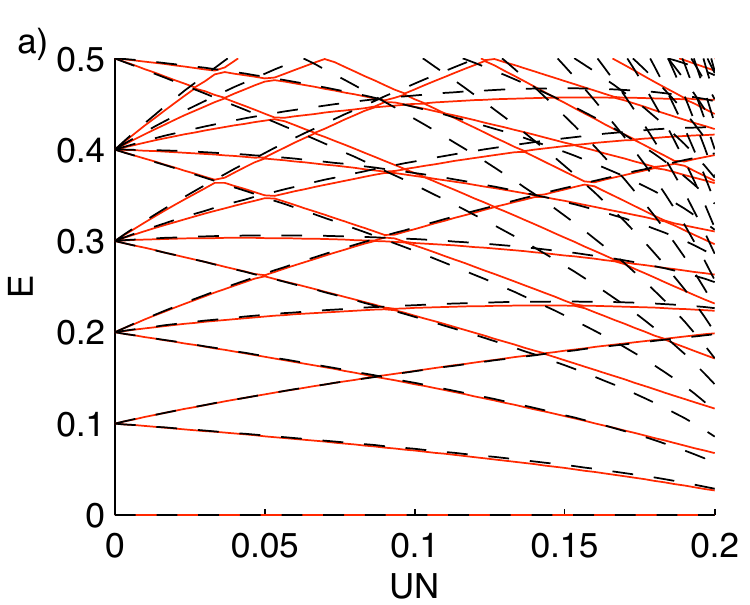}
 \includegraphics[width = 0.4\textwidth]{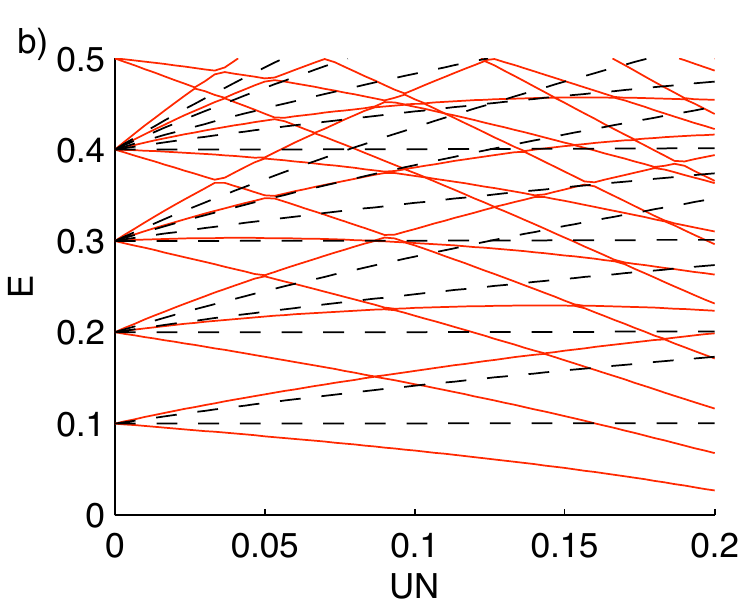}
 \caption{(Color online) Exact spectrum of the four-mode system for $N=36$, $\Omega=1$ and $\omega = 0.1$, shown is the $J=18$ band; a) solid red curve: exact diagonalization, dashed black curve: extended Bogoliubov theory; b) solid red curve: exact diagonalization, dashed black curve: standard Bogoliubov, all in units of $\hslash \Omega$}
\label{frqvgl}
\end{center}
\end{figure}

Let us now turn to the dependence of the energy levels on the interaction strength. To compare quantum rather than mean field effects for different total numbers of particles it is convenient to keep the product $UN$ constant. We may now numerically calculate the energy levels of each $J$ band and compare them with standard Bogoliubov and our extended Bogoliubov theory. Typical results are shown in Fig.~\ref{frqvgl}.
Part a) shows the numerically exact levels as solid red curves and the extended Bogoliubov theory in dashed black curves. The lowest energy level of the band has been set to zero for clarity. For moderate interaction strengths the agreement between the exact levels and the extended Bogoliubov theory is obvious, even up to higher excited states in the band. For even higher excitations, linearization clearly fails, and interactions between the quasiparticles have to be taken into account. In part b) the exact levels (again solid red) are compared to standard Bogoliubov (dashed black). The exact lower frequency spacing in the pyramid is evidently shifted below the Bogoliubov beat, as we have argued, and only approaches the Bogoliubov value in the limit $U \rightarrow 0$. For the higher of the two pyramid spacings, conversely, the standard Bogoliubov theory consistently gives a substantial underestimate. The extended Bogoliubov theory is in contrast able to capture the main features of the spectrum very well, despite being a simple linear approximation with only a two-frequency pyramid spectrum. In that sense our model is linear but non-trivial, since it includes more effects of the interactions in a resummed way via the nonlinear transformations. 

While the numerical problem becomes very hard for larger particle numbers than those presented here, experience with Bose-Hubbard systems generally indicates quite good quantum-classical correspondence, at least within the superfluid phase, so that the numerical Gross-Pitaevskii simulations presented in \cite{Strz10} should accurately represent the quantum dynamics for larger $N$. The good behavior of our theory in that regime, together with the successes shown here, therefore confirm that it is valid over a wide range of parameter space. Like any simple theory, however, it does have its limitations.

As we pointed out in our previous paper, the interactions of the josons tend to have the opposite sign to the atomic interactions. When atoms repel each other, josons attract. This leads to the possibility of dynamical instability, and therefore to a breakdown of our theory, when the joson frequency turns imaginary at a critical point. This opposite interaction of josons and atoms holds strictly in the $J \ll N$ limit, but also survives in the more general case presented in this paper, since for relatively small repulsive interactions 
\begin{equation}
  UN > \frac{\omega}{2(u+v)^2} + \mathcal{O}(\tilde J / N)
\end{equation}
the joson interaction $U_J$ becomes negative according to \eqref{abbrev}. Thus, also in this framework the joson frequency finally turns imaginary for high interactions and the model breaks down. In particular this means that the number-locked regime will not be accessible with this theory. We can see moreover that the gradual breakdown of the linear joson theory is accompanied by a series of crossings among the levels of each band. In particular even the lowest two states ultimately approach each other, and thus the low frequency goes to zero. At this point the level structure no longer consists of two linear modes, and is not be expected to be covered by a simple linear theory of this kind.

\begin{figure}[htp]
\begin{center}
 \includegraphics[width = 0.4\textwidth]{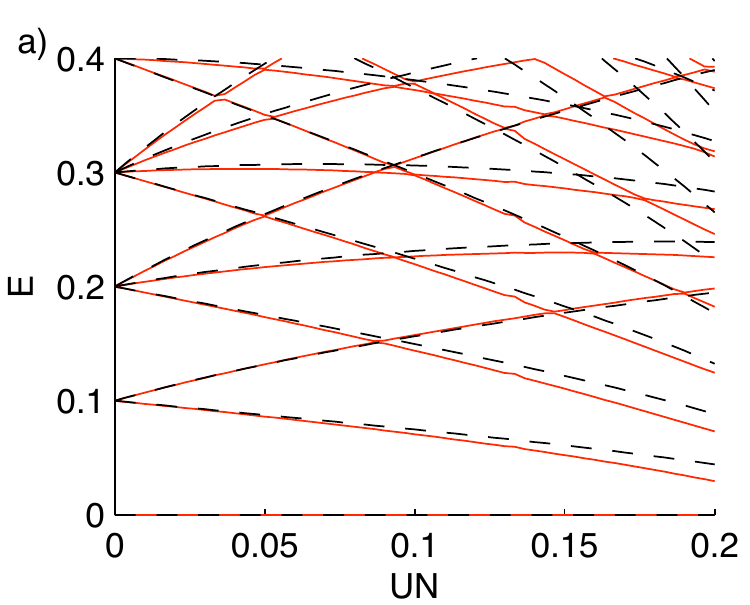}
 \includegraphics[width = 0.4\textwidth]{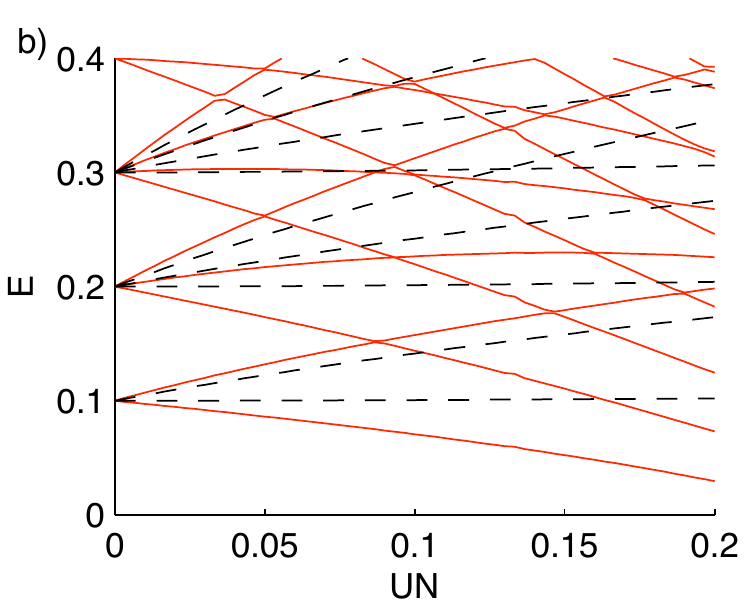}
 \caption{(Color online) Exact spectrum of the four-mode system for $N=32$, $\Omega=0.4$ and $\omega = 0.1$, shown is the $J=16$ band; a) solid red curve: exact diagonalization, dashed black curve: extended Bogoliubov theory; b) solid red curve: exact diagonalization, dashed black curve: standard Bogoliubov, all in units of $2.5\hslash \Omega$}
\label{frqvglwO}
\end{center}
\end{figure}

Another possible limitation to be explored is the reliance on the clear time scale separation. The whole system was constructed in the first place to implement a clear hirarcy of the tunneling frequencies according to $\omega / \Omega \ll 1$, to enable the adiabatic RWA to hold. We may now also numerically check the validity of the theory for larger values of the frequency ratio. The limit $\omega \rightarrow \Omega$ is rather uninteresting, since in the regime of moderate interactions the linear part of the Hamiltonian then dominates and the whole spectrum is almost linear. The limit $\Omega \rightarrow \omega$ on the contrary is more interesting. In Fig.~\ref{frqvglwO} the $J = 16$ band is shown for $N = 32$ particles and $\Omega = 0.4$ and $\omega = 0.1$.
In this case the theory still yields good agreement with the exact spectrum. For smaller values of $\Omega$ $J$ starts to be not a properly defined quantum number since fluctuations rise, which make it numerically challenging to extract the band structure correctly, especially since the bands overlap significantly in that regime. However, we can argue that the adiabatic theory does not rely on the strong adiabatic limit, but rather still holds for larger values of the frequency ratio $\omega / \Omega$. So in fact the regime of first and second Josehpson oscillations in this four-mode system is not a narrow region in parameter space, but actually the generic behavior in the regime of moderate interactions.

\section{Discussion}

In this paper we have demonstrated a powerful extension of the number-conserving Bogoliubov linearization theory which can be applied to bosonic many-body systems with time scale hierarchies. This resummed linearization conserves the number of high-frequency excitation quanta, which in exact terms is adiabatic invariant rather than strictly conserved, as well as the exactly conserved particle number. While we have demonstrated it for a very simple quantum many-body system, the general idea should be quite robust, as adiabatic methods usually are, and be extensible to more complicated systems of many kinds. The main value of the method is in providing simple quasiparticle descriptions valid in the Hilbert space neighborhoods of many excited states, rather than only of the ground state.

As well as thus demonstrating a calculational technique, we consider our study of this four-mode Bose-Hubbard system as providing a conceptually instructive toy model for an important general phenomenon: the quantum dynamics of hydrodynamic collective excitations.

Hydrodynamic collective excitations of many-body systems close to equilibrium but far from the ground state, such as sound waves in air, are usually described in terms of classical fields, such as those of temperature, velocity, and pressure. These thermodynamic quantities are in fact directly related to microscopic dynamical quantities, such as the densities of particles, momentum, or energy. The reason why these particular quantities play the main roles in low-energy effective theory is that they are conserved, and so they cannot be locally changed even by very rapid microphysics, but only by the slower processes of transport. The description of gas dynamics in terms of classical hydrodynamic collective variables such as pressure and temperature is therefore ultimately justified by a dynamical time scale hierarchy. 

But time scale hierarchies can be exploited very effectively within pure-state Hamiltonian quantum mechanics by using adiabatic theory, without assuming either statistical equilibrium or the classical limit. 

This raises the question of whether a purely quantum and Hamiltonian theory could be applied in the mesoscopic regime. 

Ordinary sound waves in air are, after all, rather simple, linear excitations. Might they not have a simple quantum description, with canonical kets expressed in a basis of quantized quasiparticle excitations, and not just as a phenomenological requantization of the classical field theory, but rigorously derived from many-body first principles? This question may be entirely academic for macroscopic systems with a huge number of degrees of freedom, but is important in the mesoscopic regime, where such theories might be derived directly from the many-body Hamiltonian, as in the simple example of our four-mode BH system. Thus, this issue is very important for understanding the mesoscopic interface between quantum mechanics and thermodynamics, since here the new approach can directly be compared to the standard thermodynamical treatment. And of course the interface between quantum mechanics and thermodynamics in general is an important frontier of fundamental physics, which with the advent of nanotechnology and quantum information processing may even turn out to be of practical significance.

The quantum theory of hydrodynamics can be approached by considering a volume of gas to consist of many adjacent sub-volumes, such that the free motion of atoms across the arbitrarily defined sub-volume boundaries provides a coupling between the sub-volumes. Motion and collisions within each sub-volume are then considered as internal dynamics for the sub-volume, considered as a dynamical sub-system of the whole gas. By definition, then, each sub-system's internal dynamics conserves the sub-system's total energy, momentum, and particle number. If the subvolumes are large enough, the rate of proportional change in these quantities, due to particles moving between sub-volumes, must be low. The effective low-frequency description of the problem therefore takes these slowly changing quantities as its elementary degrees of freedom.

This means, in particular, that the local sub-volume energy is effectively treated in much the same way as the sub-volume particle number. The local energy might almost be a second kind of fluid, mixed together with the fluid of particles. Most of this energy is usually thermal, and in this sense the kinetic theory of heat reduces, as a hydrodynamic theory at low frequencies, to something very much like the previous theory which it replaced -- the caloric theory of heat developed in the 18th century by Lavoisier and Laplace, according to which heat was considered a material fluid analogous to electric charge.

The effective behavior of energy as a dynamical degree of freedom is in fact familiar even in dynamical systems much simpler than an interacting gas. To observe a pair of weakly coupled pendulums in underdamped motion, and see how the swing amplitude oscillates slowly back and forth from one pendulum to the other, is to notice that the energy distribution between the two pendulums behaves very much like a pendulum itself.

The contribution we have made in this paper is to study a quantum many-body system that is almost as simple as those pendulums, and show how the low-frequency description of local energy as an effective second fluid can be implemented quantum mechanically, entirely within a canonical Hamiltonian closed system. In effect we have simplified the hydrodynamic construction of sub-volumes to the minimal case of just two neighboring sub-volumes, and enforced relatively slow motion between them by requiring it to proceed by slow tunneling. We have then further simplified each sub-volume by idealizing them as two-site bosonic lattice gases, modeling their internal sub-volume dynamics in quantum Josephson form. As a toy model for the hydrodynamic problem, then, our system is clearly crude to the point of absolute minimality, but does manage to capture some qualitatively crucial features of the real phenomenon. Because we have gone beyond the trivial case of non-interacting particles, moreover, we have identified some non-trivial interaction effects. 

In our previous paper on this system, we showed that the (quasi-)particles of the `heat fluid,' which we have called `josons', tend to attract each other if the fundamental particles repel one another, and vice versa. We then showed in consequence of this that for repulsively interacting atoms, the attractive interaction of josons makes the transport of `heat' between our two subsystems slower than one would anticipate from a naive application of the coupled pendulum picture. We have now shown how this picture persists into the fully quantum regime. In particular we have demonstrated that an extended Bogoliubov theory, based on treating the joson number $J$ on the same footing as the atom number $N$, can yield good approximations for the exact many-body energy levels within each $(N,J)$ band. There is a somewhat dizzying but pleasing symmetry in the way that the successive Bogoliubov transformations absorb subtler forms of interaction into new forms on non-interacting quasi-particle. Just as the first Bogoliubov transformation absorbs nonlinear corrections to Rabi oscillations into the frequency-renormalized Josephson oscillations, so the second Bogoliubov transformation absorbs nonlinear corrections to the linear Bogoliubov theory into frequency-renormalized second Josephson oscillations. This is, after all, what hydrodynamics is all about: the emergence of simple collective modes from highly nonlinear microphysics.

While fully hydrodynamic and thermodynamic phenomena such as irreversibility and turbulence are obviously far beyond the scope of our minimal toy model, nonetheless we can already see what may perhaps be some first, faint indications of how they emerge from Hamiltonian closed-system quantum mechanics. Our numerical plots of exact energy levels show a number of avoided crossings, which our linear extended Bogoliubov theory naturally does not recognize, even where it otherwise follows the exact level curves very well. In systems with many more degrees of freedom, one can anticipate that such avoided crossings may perhaps proliferate until the free-quasiparticle theory is thoroughly eroded. Bose-Hubbard systems such as ours tend in general to be dynamically chaotic, and the application of adiabatic theory such as ours in the presence of chaos presents numerous difficulties \cite{Kolo07a,Kolo07b}. With further research, the nonlinear quantum dynamics of simple many-body systems with time scale hierarchies may offer further insights into important basic problems.


\end{document}